\documentstyle[aps,preprint,prl,epsf]{revtex}

\newcommand{\bn}{\mbox{\boldmath{$n$}}}
\newcommand{\bxi}{\mbox{\boldmath{$\xi$}}}
\newcommand{\bx}{\mbox{\boldmath{$x$}}}
\newcommand{\by}{\mbox{\boldmath{$y$}}}
\newcommand{\bS}{\mbox{\boldmath{$S$}}}
\newcommand{\bz}{\mbox{\boldmath{$z$}}}
\newcommand{\bzeta}{\mbox{\boldmath{$\zeta$}}}
\newcommand{\bJ}{\mbox{\boldmath{$J$}}}

\begin{document}

\title{Tighter Decoding Reliability Bound for Gallager's 
Error-Correcting Code}
\author{Yoshiyuki~Kabashima$^{1}$, Naoya Sazuka$^{1}$, Kazutaka Nakamura$^{1}$
and David~Saad$^{2}$}
\address{$^{1}$ Department of Computational Intelligence and Systems Science,
Tokyo Institute of Technology, Yokohama 2268502, Japan.  \\
$^{2}$The Neural Computing Research Group, Aston
University, Birmingham B4 7ET, UK.}
\maketitle
\begin{abstract}

Statistical physics is employed to evaluate the performance of
error-correcting codes in the case of finite message length for an
ensemble of Gallager's error correcting codes.  We follow Gallager's
approach of upper-bounding the average decoding error rate, but invoke
the replica method to reproduce the tightest general bound to date,
and to improve on the most accurate zero-error noise level threshold
reported in the literature. The relation between the methods used and
those presented in the information theory literature are explored.

\end{abstract}


Many of the problems addressed in the Information Theory (IT)
literature show great similarity to those treated in statistical
physics. One of the main areas where these links are particularly
strong is that of digital communication and coding theory; these links
have been recently examined in the area of Low Density Parity Check
(LPDC)~\cite{Sourlas,us_PRL} and turbo~\cite{turbo_sourlas}
error-correcting codes.  It is only natural to expect that some
relations between the analytical methods used in the two disciplines
will emerge, and that advances in one could be employed to improve
results in the other.  In this Letter we focus on such an example. We
utilize the replica method of statistical physics to assess the
performance of Gallager's error correcting code in the case of finite
message length, generalizing an established method in the IT
community.  The analysis reproduces the tightest general bound to
date, but more importantly, it provides exact results to specific code
constructions.

Error correcting codes play a vital role in facilitating reliable data
transmission, ranging from cellular communication to data storage on
magnetic media. In a general scenario, the $N$ dimensional Boolean
message $\bxi\in\{0,1\}^N$ is encoded to the $M(>N)$ dimensional
Boolean vector $\bz_0$, and transmitted via a noisy channel, which is
taken here to be a Binary Symmetric Channel (BSC) characterized by
flip probability $p$ per bit; other transmission channels may also be
examined within a similar framework. At the other end of the channel,
the corrupted codeword is decoded utilizing the structured codeword
redundancy.

The block error rate $P_E$, defined as the probability for a decoding
error, serves as a performance measure for the success of the coding
method. In his seminal work~\cite{Shannon}, Shannon showed that the
error rate can vanish for code rates $R$ below the channel capacity in
the limit $N, M \to \infty$; in the case of the BSC and unbiased
messages $ R \!=\! N/M \!<\!  1\!-\!H_2(p), $ where $H_2(p)\! =
\!-p\log_2 p-(1-p)\log_2(1-p)$.  The upper bound, for infinitely long
messages, is often termed {\em Shannon's limit} to the error
correcting ability. Evaluating $P_E$ for practical codes of finite
length became one of central topics in IT.

For maximum likelihood (ML) decoding where the most probable message
given the possibly corrupted codeword defines the message
estimate, it is believed that $P_E$ of the best code scales as $\exp
[-ME(R)]$.  The non-negative exponent $E(R)$ is termed {\em
reliability function} (RF); it becomes positive below the channel
capacity defining the sensitivity of the optimal error rate to the
message length, complementing Shannon's result.

Unfortunately, assessing the RF directly is generally difficult. 
Instead, Gallager's powerful method~\cite{Gallager} bounds $E(R)$ 
from the below utilizing the inequality 
\begin{equation}
P_E \le \mathop{\rm Tr}_{\{\by, \bx\}} P^{\frac{1}{1+\rho}}(\by,\bx)
\left ( \mathop{\rm Tr}_{\{\bx^\prime \ne \bx \}} P^{\frac{1}{1+\rho}}(\by,
\bx^\prime) \right )^\rho \ , 
\label{eq:gallager}
\end{equation}
which holds for any arbitrary ML estimation, inferring a binary vector
$\bx$ after observing a vector $\by$, and a positive variable $\rho\!
>\! 0$.

The average error rate $\bar{P}_E$ for a certain ensemble of codes is
greater than the ensemble minimum.  Therefore, averaging the RHS of
Eq.(\ref{eq:gallager}) over the ensemble, one obtains an upper-bound
to the minimum error rate that scales exponentially with $M$ for large
but finite $N$ and $M$, $\exp[-M E_{av}(\rho,R)]$; the exponent
$E_{av}(\rho,R)$ serves as a {\em lower-bound} of $E(R)$.  One can
tighten the lower bound by maximizing $E_{av}(\rho,R)$ with respect to
$\rho\! >\! 0$.

Evaluating $E_{av}(\rho,R)$ is also difficult (except for $\rho
\!\in\! I\!\!N$). The strategy used by Gallager~\cite{Gallager} is to
further upper-bound the RHS of Eq.(\ref{eq:gallager}) utilizing
Jensen's inequality $\langle x^\rho \rangle \!\le\! \langle x
\rangle^\rho$, which holds for any $0 \!\le\! \rho \!\le\! 1$ with
respect to the expectation over any arbitrary distribution of a
positive variable.  The added inequality presumably makes the bound
looser. It is therefore surprising that maximizing the exponent with
respect to $\rho \in [0,1]$ in the ensemble of all random codes having
the same rate $R$, which results in the {\em random coding exponent}
$E_r(R)$, provides an exact evaluation of the RF for high $R$ values.

However, the bound by $E_r(R)$ becomes loose once the optimal value of
$\rho$ reaches the upper limit of the interval, i.e., $\rho\! = \!1$
(corresponding to Bhattacharyya's bound).  It is not clear whether
Jensen's inequality or Gallager's inequality~(\ref{eq:gallager}) is
responsible for this breakdown.  Moreover, it is unclear how to devise
a similar method for deriving bounds for other (non-random) codes, a
question of high practical significance.

In this Letter we demonstrate how the methods of statistical physics
may be employed to obtain tighter bounds for specific codes. This is
carried out by a direct evaluation of $E_{av}(\rho,R)$ for the
ensemble of Gallager error-correcting codes~\cite{Gallager_code}.
This (linear) code was rediscovered only recently~\cite{MacKay},
showing outstanding performance, competitive to other state-of-the-art
techniques. It is characterized by a randomly generated $(M-N) \times
M$ Boolean sparse parity check matrix $H$, composed of $K$ and $C \
(\ge 3)$ non-zero (unit) elements per row and column,
respectively. Encoding the message vector $\bxi$, is carried out using
the $M \times N$ generating matrix $G^T$, satisfying the condition
$HG^T\! = \!0$, where $\bz_0\! = \!G^T \bxi \ (\mbox{mod $2$})$.  The
$M$ bit codeword $\bz_0$ is transmitted via a noisy channel, BSC in
the current analysis; the corrupted vector $\bz\! = \!\bz_0+\bzeta \
(\mbox{mod $2$})$ is received at the other end, where
$\bzeta\!\in\!\{0,1\}^M$ represents a noise vector with an independent
probability $p$ per bit of having a value 1. Decoding is carried out
by multiplying $\bz$ by the parity check matrix $H$, to obtain the
syndrome vector $\bJ\! = \!H \bz \! = \!  H(G^T\bxi+\bzeta)\! =
\!H\bzeta \ (\mbox{mod $2$})$, and to find the most probable solution
to the parity check equation
$
H \bn \! = \! \bJ  \ (\mbox{mod $2$}) \ , 
$
for estimating the true noise vector $\bzeta$. One retrieves the original
message using the equation $G^T \bS \! = \! \bz\! -\! \bn \ (\mbox{mod
$2$})$; $\bS$ to estimate of the original message.

To facilitate the analysis we map the Boolean $(0,1)$ variables onto
the binary $(\pm 1)$ representation. The binary vectors $\bn$ and
$\bJ$, represent the noise estimate and syndrome vectors
respectively; the latter is generated by taking products of the
relevant noise bits $J_\mu \! = \! \zeta_{i_{1 \mu}}.. \zeta_{i_{K
\mu}}$, where the indices $i_{1 \mu},.., i_{K \mu}$ correspond to the
nonzero elements in row $\mu$ of the parity check matrix~$H$.

The similarity between error-correcting codes and physical systems was
first pointed out by Sourlas~\cite{Sourlas}, mapping a simple
Boolean code onto Ising spin models with multi-spin interactions.  We
recently extended his work to more practical parity check
codes~\cite{us_PRL}. We employ a similar formulation using the Hamiltonian
\begin{equation} 
{\cal H}(\bn;\bJ) \! = \! \gamma \sum_{\cal G} D_{\cal G} \
\delta \ \left (J_{\cal G};-\prod_{i \in {\cal G}} n_i \right ) -F
\sum_{i=1}^M n_i \ , 
\label{eq:hamiltonian} 
\end{equation} 
to evaluate the joint probability for $\bJ$ and  $\bn$
\begin{eqnarray} 
P\left ( \bJ,\bn \right )  \! = \!
\lim_{\gamma \to \infty} \frac{\exp[-\beta {\cal H}(\bn;\bJ) ]}{\left
( 2 \cosh F \right )^M} \ .  
\label{eq:joint} 
\end{eqnarray} 
Here, ${\cal G}\! \equiv\! \left \langle i_1,..,i_K \right \rangle$
runs over all combinations of $K$ indices out of $M$;
$J_{\cal G} \!\equiv \!\prod_{i \in {\cal G}} \zeta_i$ and the sparse
tensor $D_{\cal G}$ becomes non-zero (unit) only when all indices in
$\cal G$ correspond to non-zero (unit) elements in a certain row of
the parity check matrix $H$.  Taking $\gamma \!\to\! \infty$ enforces the
parity check equation.
The additive field $F\!
= \!(1/2)\ln \left [ (1\!-\!p)/p \right ]$ corresponds to the true prior
probability in the Bayesian framework, reflecting the flip
rate $p$.  The inverse temperature $\beta$ is introduced to
emphasize the link with the statistical mechanics formulation and is
generally fixed to $\beta\! = \!1$ unless specified otherwise.

One can then use (\ref{eq:joint}) to evaluate $\bar{P}_E$  from
(\ref{eq:gallager}) by calculating the bound without invoking
Jensen's inequality.  The first part of the Hamiltonian
(\ref{eq:hamiltonian}) is invariant under gauge transformations of the
form $n_{i} \!\rightarrow\!  n_{i} \zeta_{i}$, and $J_{\cal G}
\!\rightarrow\! J_{\cal G}\prod_{i \in {\cal G}} \zeta_i \!  = \! 1$,
which decouple the correlation between the dynamical vector $\bn$ and
the true noise $\bzeta$. Rewriting the Hamiltonian one obtains a
similar expression to Eq.~(\ref{eq:hamiltonian}) apart from the last
term on the right which become $ F \sum_{i} \zeta_i n_{i} $.

Quenched averages over the ensemble of codes is carried out with
respect to the current random selection of the sparse tensor $D$ and
the noise vector, which eventually results in a similar procedure to
the replica method in statistical mechanics.  This gives rise to a set
of order parameters $ q_{\alpha, \beta, \ldots, \gamma} \! = \!
\frac{1}{M} \sum_{i\! = \!1}^{M} Z_{i} \ n_{i}^{\alpha} \
n_{i}^{\beta}...n_{i}^{\gamma} \ , $ where $\alpha$, $\beta \ldots$
represent replica indices, and the variable $Z_{i}$ comes from
enforcing the restriction of $C$ and $L$ connections per index
respectively as in~\cite{us_PRL}.  This interesting similarity between
Gallager's method and the replica method has been pointed out by Iba
in \cite{Iba2}.

To proceed further one has to make an assumption about the order
parameter symmetry. As a first approximation we assume replica
symmetry (RS) in the following order parameters and the related
conjugate variables
\begin{eqnarray}
\label{eq:order_parameters_RS}
q_{\alpha, \beta,.., \gamma} \! = \! q \int d x \ \pi(x) x^{l} \ ,  \ 
\widehat{q}_{\alpha, \beta,.., \gamma} \! = \! 
\widehat{q} \int d \hat{x} \ \widehat{\pi}(\hat{x}) \ \hat{x}^{l} \ ,
\end{eqnarray}
where $l$ is the number of replica indices, $q$ and $\widehat{q}$ are
normalization variables ($\pi(x)$ and $\widehat{\pi}(\hat{x})$ are
 probability distributions).  Unspecified integrals are over
the range $[-1,+1]$. 

Originally, the summation $\mathop{\rm Tr}_{\{ \bn \ne \bzeta \}} (
\cdot )$ excludes the case of $\bn \ne \bzeta$; however, it can be
shown that in the limit of large $M$ this becomes identical to the
full summation in the non-ferromagnetic phase, where $\pi(x) \ne
\delta (x-1)$ and $\widehat{\pi}(\widehat{x}) \ne \delta
(\widehat{x}-1)$.  Then, one obtains the expression
\begin{eqnarray}
E_{av}(\rho,R)&\! = \!& -\frac{1}{M} \ln  \left [
\left \langle 
\mathop{\rm Tr}_{\{\bJ, \bzeta\}} P^{\frac{1}{1+\rho}}(\bJ,\bzeta)
\left ( \mathop{\rm Tr}_{\{\bn \ne \bzeta \}} P^{\frac{1}{1+\rho}}(\bJ,
\bn) 
\right )^\rho
\right \rangle_D \right ] \cr 
&\! = \!&
\ln \left ( 2 \cosh F \right ) - 
\ln \left ( 2 \cosh \left ( \frac{F}{1+\rho} \right ) \right )-
\frac{1}{M} 
\ln \left \langle {\cal Z}_{\rm NF}^\rho\left (\bzeta,D;\frac{F}{1+\rho} 
\right ) \right \rangle_{\bzeta \mid \frac{F}{1+\rho}, D} \ , 
\label{eq:E_rho}
\end{eqnarray}
where ${\cal Z}_{\rm NF} (\bzeta,D;\frac{F}{1+\rho})$ denotes the
partition function $\mathop{\rm Tr}_{\bn} \lim_{\gamma \to \infty}
\exp [ -\beta {\cal H} ]$ in the non-ferromagnetic phase for a system
with an effective additive field $F/(1+\rho)$.  Averages $\left
\langle \cdot \right \rangle_{\bzeta \mid \frac{F}{1+\rho},D}$ are
over the distribution
$P(\bzeta;\frac{F}{1+\rho})=\exp[\frac{F}{1+\rho}\sum_{i=1}^M
\zeta_i]/ \left (2 \cosh \left (\frac{F}{1+\rho} \right) \right )^M$
and the uniform distribution of $D$. Extremizing $\left \langle {\cal
Z}_{\rm NF}^\rho\left (\bzeta,D;\frac{F}{1+\rho} \right ) \right
%david - minor change
\rangle_{\bzeta\mid\frac{F}{1+\rho}, D}$ with respect to the order
parameters $q, \ \widehat{q}, \ \pi(\cdot)$ and $\widehat{\pi}(\cdot)$,
under the replica symmetry ansatz (\ref{eq:order_parameters_RS}), one
obtains for the final term in (\ref{eq:E_rho})
\begin{eqnarray}
& & \frac{1}{M} 
\ln \left \langle {\cal Z}_{\rm NF}^\rho\left (\bzeta,D;\frac{F}{1+\rho} 
\right ) \right \rangle_{\bzeta\mid\frac{F}{1+\rho}, D} \! = \!
\mathop{\mbox{\rm Ext}^*}_{ \{ q,\widehat{q}, 
\pi(\cdot),\widehat{\pi}(\cdot) \} }
\left \{
\frac{C \ q^K}{K} \int \prod_{i\! = \!1}^K d x_i \pi(x_i) 
\left 
(\frac{ 1+ \prod_{i\! = \!1}^K x_i}{2} \right )^\rho 
\right . \cr 
&+& \ln \left [
\int \prod_{\mu\! = \!1}^C d \widehat{x}_{\mu} \ \widehat{\pi} (\widehat{x}_{\mu})
\left \langle 
\left ( e^{\frac{F}{1+\rho} \zeta } 
\prod_{\mu\! = \!1}^C \left ( \frac{1+\widehat{x}_\mu}{2} \right )
\right . \right . \right .
+ \left . \left . \left . e^{-\frac{F}{1+\rho} \zeta }
\prod_{\mu\! = \!1}^C \left ( \frac{1-\widehat{x}_\mu}{2} \right )
\right )^\rho \right \rangle_{\zeta\mid\frac{F}{1+\rho}}
\right ]  \cr
&+& \left . C \ln \widehat{q} - 
C q \widehat{q} \int dx \ d \widehat{x} \ \pi(x) \ \widehat{\pi}(\widehat{x}) 
\left (\frac{ 1+ x \widehat{x}}{2} \right )^\rho 
-\left ( \frac{C}{K}-C \right )
\right \}, 
\label{eq:Z_rho}
\end{eqnarray}
where $\mbox{\rm Ext}^*$ denotes extremization which excludes the
ferro-magnetic solution and $\left \langle \cdot \right
\rangle_{\zeta\mid\frac{F}{1+\rho}} $ is over
$P(\bzeta;\frac{F}{1+\rho})$.

Before proceeding any further, we would like to mention some general
properties of $E_{av}(\rho,R)$.  From Eqs.~(\ref{eq:E_rho}) and
(\ref{eq:Z_rho}), it can be shown that $\lim_{\rho \to 0}
E_{av}(\rho,R) \! = \!  0$ and $\partial^2 E_{av}(\rho,R)/\partial
\rho^2\!<\!0$.  This implies that $\mathop{\rm Max}_{\rho > 0}
E_{av}(\rho,R)$, becomes positive if and only if $\left. \partial
E_{av}(\rho,R)/\partial \rho \right|_{\rho = 0} > 0$, for which
$\lim_{M\to\infty} \bar{P}_E \! = \! 0$ holds.
Therefore, the zero error threshold, defined as the critical flip rate
below which the average error rate vanishes as $M\to\infty$, is
obtained by the condition $ \partial E_{av}(\rho,R)/ \partial \rho \!
= \!0$.  From (\ref{eq:E_rho}), this becomes
\begin{equation}
F \tanh F - \frac{1}{M} \left \langle \ln {\cal Z}_{\rm NF} \left
(\bzeta,D;F \right ) \right \rangle_{\bzeta \mid F,D} \! = \!0.
\label{eq:threshold}
\end{equation}
The second term is the averaged free energy for the Hamiltonian
(\ref{eq:hamiltonian}) with respect to the quenched randomness
$\bzeta$ and $D$, in the non-ferromagnetic phase. Employing the
ferromagnetic gauge~\cite{Nishimori} one obtains the following
expression for the ferromagnetic free energy (where $\bar{P}_E\! =
\!0$): $(1/M) \left \langle \ln {\cal Z}_{\rm F} \left (\bzeta,D;F
\right ) \right \rangle_{\bzeta \mid F,D} \! = \!F \tanh F$.  Since
the correct prior information about the flip rate $p$ is used in the
calculation, these two free energies are actually obtained in
Nishimori's {\em finite} decoding temperature ($\beta\! =
\!1$)~\cite{Sourlas,Rujan,Nishimori,Iba} for which the bit error
probability is minimized.  By satisfying (\ref{eq:threshold}), the
zero error threshold for ML decoding, which corresponds to the zero
temperature limit $(\beta \! \rightarrow \!
\infty)$~\cite{Sourlas,Iba}, is determined by the phase boundary
between the ferromagnetic and non-ferromagnetic phases at $\beta\! =
\!1$.

Using the ferromagnetic gauge provides insight into the physical
properties of the system. As the internal energy per bit in the
non-ferromagnetic system is $-F \tanh F$ under Nishimori's condition,
Eq.~(\ref{eq:threshold}) implies that the entropy of the
non-ferromagnetic phase vanishes at the phase boundary for $\beta\! =
\!1$, suggesting that this phase exhibits a replica symmetry breaking
(RSB) at lower temperatures in general, and at $\beta \! \rightarrow
\! \infty$ in particular.  In this sense, the zero-error threshold
prediction obtained from Gallager's method and ML decoding, is
surprising as it provides information about the ferro/non-ferro phase
boundary at $\beta \! \rightarrow\! \infty$ which is not easily 
obtained via the methods of statistical physics due to RSB effects. This
argument can be extended to the case of general $\beta \ge 1$, as will
be presented elsewhere.

An analytical expression to $E_{av}(\rho,R)$ can be obtained in the limit
$K,C \!\to\! \infty$, keeping the code rate $R\! = \!1\!-\!C/K$
finite; for the non-ferromagnetic solution one then obtains $ q\! =
\!2^{\rho/K}, \ 
\widehat{q}\! = \!2^{\rho(1\!-\! 1/\!K)}, \ \pi(x)\! =
\!\delta(x)$ and $\widehat{\pi}(\widehat{x})\! = \!(1/2) (1\!+\!\tanh F)
% yoshiyuki corrected 
%\delta(\widehat{x}\!-\!F) \!+\! (1/2) (1\!-\!\tanh F)
%\delta(\widehat{x}\!+\!F).  $
\delta(\widehat{x}\!-\! \tanh F) \!+\! (1/2) (1\!-\!\tanh F)
\delta(\widehat{x}\!+\! \tanh F).  $
Using Eqs.~(\ref{eq:E_rho}) and (\ref{eq:Z_rho}), one obtains the
explicit expression
$E_{av}(\rho, R)\!=\! \ln 2 \cosh F \!-\!
(1\!+\! \rho ) \ln \left ( 2 \cosh \frac{F}{1\!+\!\rho} \right ) \!+\!
\rho (1\!-\!R) \ln 2 $.  
In addition, there exists another solution for $\rho \ge 1$, 
$q=2^{1/K}, \ \widehat{q}=2^{1-1/K}, \
\ \pi(x)\! =
\! (1/2) \delta(x-1) \! + \! (1/2) \delta(x \! + \!1)$ and 
$\widehat{\pi}(\widehat{x})\! = \! (1/2) \delta(\widehat{x}\!-\!1) 
\!+\! (1/2) \delta(\widehat{x}\!+\!1) $
providing 
$E_{av}(\rho, R) \!=\! \ln 2 \cosh F \! - \! 
\ln \left (2 \cosh F \! + \! 2 \cosh \left (\frac{1-\rho}{1+\rho}F \right )
\right ) \! +\! (1-R)\ln2 $.
Employing a method similar to that in \cite{Mottishaw,turbo_sourlas}, 
it can be shown that both RS solutions are locally stable 
against perturbations to the replica symmetric solution. 

The relation between $E_{av}(\rho,R)$ and the entropy of non-ferromagnetic 
solutions ${\cal S}_{\rm NF}$
\[
\frac{\partial E_{av}(\rho,R)}{\partial \rho} \! = \! 
-\frac{\langle {\cal Z}_{\rm NF}^\rho\left (\bzeta,D;\frac{F}{1+\rho} 
\right ) {\cal S}_{\rm NF}\left (\bzeta,D;\frac{F}{1+\rho} \right ) 
\rangle_{\bzeta\mid\frac{F}{1+\rho}, D}}{
\langle {\cal Z}_{\rm NF}^\rho\left (\bzeta,D;\frac{F}{1+\rho} 
\right ) 
\rangle_{\bzeta\mid\frac{F}{1+\rho}, D}} \ , 
\]
suggests another type of RSB, indicated by the negative entropy.  This
implies that the entropy of the non-ferromagnetic RS solutions
vanishes at $\rho=\rho^*(R)$ which maximizes $E_{av}(\rho,R)$; and the
tightest lower bound of $E(R)$ is therefore obtained at the RSB
transition, which can be calculated from the locally stable RS
solutions.

Solving the maximization problem one obtains
\begin{eqnarray} 
\mathop{\rm Max}_{\rho >0}  E_{av}(\rho,R) 
\! = \! \left \{
\begin{array}{ll}
\ln 2 \cosh F \!-\! (1\!-\!R) \ln 2 & \ \ F \! \ge \! 2 F^*(R) \\
\!-\! \ln \left (2 \cosh F\!+\! 2 \right ) \ , & \\ \ln 2 \cosh F
\!-\! (1\!-\!R) \ln 2 & \ \ 2 F^*(R) \! \ge \! F \! \ge \! F^*(R) \\
\!-\!F \tanh F^*(R) \ , & \\ 0 \ , & \mbox{otherwise}
\end{array}
\right . 
\label{eq:E_R}
\end{eqnarray}
where $F^*(R)$ is the solution of the equation $\ln 2 \cosh F^* \!-\!
F^* \tanh F^* \!-\!(1\!-\!R) \ln 2 \! = \!0$. 
The position of the maximum is given as 
$\rho^*(R)\! = 1$ for $F \! \ge \! 2 F^*(R) $, \ $\!F/F^*(R)\!-\!1$ for
$2 F^*(R) \! \ge \! F \! \ge \! F^*(R)$ and $0$, otherwise.
Using the relation between $F$ and $p$, this indicates that
$E(R)$ becomes positive if and only if $R \!<\! 1\!-\!H_2(p)$, 
which corresponds to Shannon's limit.

Equation~(\ref{eq:E_R}) is identical to the random coding exponent
$E_{r}(R)$ obtained in the IT literature~\cite{Gallager}, although one
should emphasize the main differences between the two approaches: a)
Strating from Gallager's inequality (\ref{eq:gallager}) we directly
average over the ensemble while the $E_r(R)$ result is obtained by
invoking Jensen's inequality. b) Our result is obtained for an
ensemble of a specific code.

With some hindsight, this is not very surprising as Gallager codes
become similar to random codes in the limit $K, C \to
\infty$~\cite{MacKay,us_PRL}; this also implies that using Jensen's
% yoshiyuki corrected 
%inequality does produce a looser bound as initially thought.
inequality does not produce a looser bound as initially thought.

To get a tighter bound for low $R$ values we employ a refined
inequality, upper-bounding the ensemble minimum of $P_E$ by $\left
\langle \left (\mathop{\rm Tr}_{\{\bJ,\bzeta\}}
P^{\frac{1}{1+\rho}}(\bJ,\bzeta) \left (\mathop{\rm Tr}_{\{\bJ,\bn \ne
\bzeta \}}P^{\frac{1}{1+\rho}} (\bJ,\bn) \right )^\rho \right )^m
\right \rangle_D^{\frac{1}{m}}$ ($\rho>0, \ m >0$), as in
(\ref{eq:gallager}).  A similar calculation along the lines described
here (details will be shown elsewhere) provides the {\em expurgated
exponent} bound~\cite{Gallager} result for low $R$ values (see Fig.1);
this links our results to the best bounds reported in the IT
litereture to date.

Without trivializing the results obtained in the case of $K, C \to
\infty$, the main achievement of our approach is the ability to
investigate analytically the performance of Gallager (or similar)
codes of finite $K$ and $C$.  To demonstrate the accuracy of the
bounds obtained we examine the case of $K\! = \!6$ and $C\! = \!3$. We
numerically evaluated $E_{av}(\rho,R)$ (\ref{eq:E_rho}) for $p\! =
\!0.0915$, a recent highly accurate estimate of the error threshold
for this parameter~\cite{Aji}, and for $p\! = \!0.0990$, which is the
threshold predicted by our analysis.  The numerical results were
obtained by approximating $\pi(\cdot)$ and $\widehat{\pi}(\cdot)$
using $10^6$ dimensional vectors and iterating the saddle point
equations until convergence.  The results are shown in the inset; they
indicate that $\mathop{\rm Max}_{\rho \ge 0 }{E_{av}(\rho,R)} \simeq
1.0 \times 10^{-4} > 0$ for $p\! = \!0.0915$ while $E_{av}(\rho,R)$ is
maximized (to zero) in the vicinity of $\rho \! = \! 0$ for $p\! =
\!0.0990$, suggesting a tighter estimate for the error threshold than
those reported in the IT literature.

In summary, we have developed a method to tightly upper-bound the
dependence of the decoding error rate on the message length for
Gallager codes. In the limit of infinite connectivity our result
collapses onto the best general random coding exponents reported in
the IT literatures, the {\em random coding exponent} and the {\em
expurgated exponent} for high and low $R$ values respectively.  The
method provides one of the only tools available for examining codes of
finite connectivity; and predicts the tightest estimate of the zero
error noise level threshold to date for Gallager codes. It can be
easily extended to investigate other linear codes of a similar type
and is clearly of high practical significance.

We demonstrated how the methods of statistical
physics may complement and improve results obtained in the IT
literature.  These methods are applicable to a broad range of
problems, especially within the sub-field of coding, and may be
instrumental in improving existing results; some of these studies are
already under way.

{\small
\vspace*{0.2in} {\bf Acknowledgement } We acknowledge 
support from the JSPS-RFTF program (YK),  EPSRC 
(GR/N00562) and the Royal Society (DS).  YK would like to thank
Y.~Iba for kindly showing him an unpublished manuscript~\cite{Iba2}
and D.J.C.~MacKay for informing us of~\cite{Aji} prior to publication.  }

% FIGURE 1
% Error Exponent 
%\vspace*{0.2in}
\begin{figure}
\begin{center}
\epsfxsize=140mm  
\epsfbox{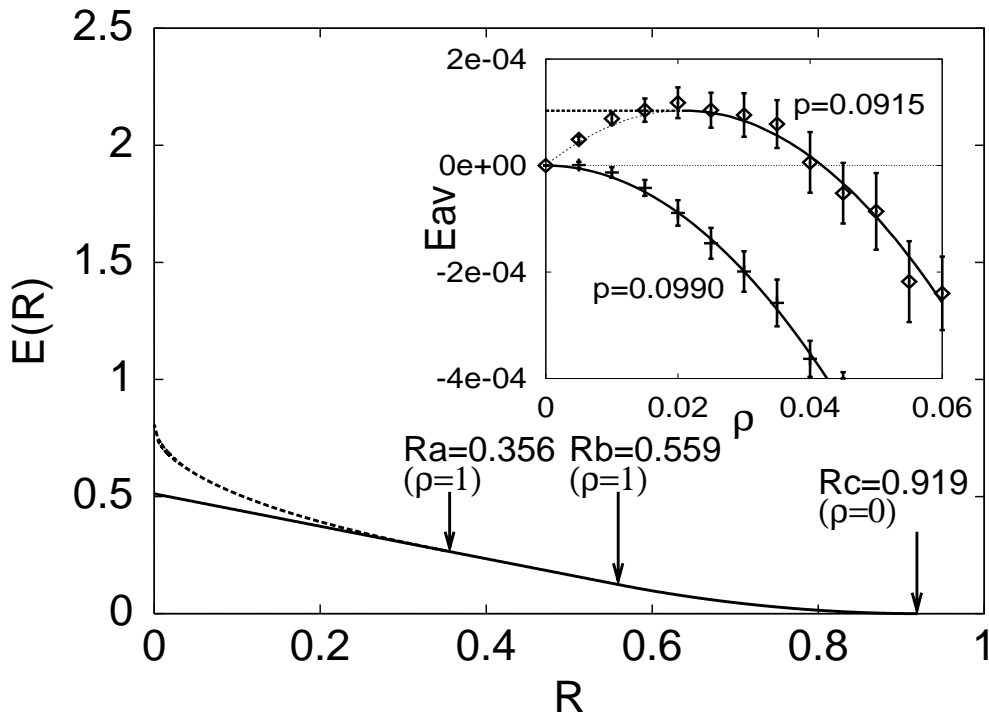}
\caption{Lower-bounds on the reliability exponent $E(R)$ obtained for
$p\! = \!0.01$ in the limit $K,C \!\to\! \infty$.  Our method produces
the same result as the random coding exponent $E_r(R)$ (solid line)
which provides an excellent bound for $R \!>\! R_b $. For low $R\! <
\!R_a $ values the bound becomes loose, and a better result (dashed
line), identical to the expurgated exponent bound, is obtained (see
text) by employing a refined inequality in (\ref{eq:gallager}).  Inset
- The exponent $E_{av}(\rho,R)$ obtained numerically for a choice of
finite parameters $K\! = \!6$ and $C\! = \!3$ ($R\! = \!1/2$). Symbols
and and standard deviations are computed using 50 numerical
solutions. Curves are obtained via a quadratic fit.  For $p\! =
\!0.0915$, $\rho^*(R) \! \simeq \!0.02$, suggesting that this flip
rate is still below the threshold.  Left of the peak, the RS solution
(thin broken curve) is unstable.  For $p\! = \!0.0990$, our predicted
threshold, the maximum $E_{av}(\rho,R) \!\simeq\! 0$ is obtained at
$\rho\! \simeq\! 0$, implying that this is the correct threshold.}
\end{center}
\label{fig:E_rho} 
\end{figure}
\end{document}